\documentclass[DIV=calc, paper=a4, fontsize=10pt, twocolumn]{scrartcl}	 

\usepackage[english]{babel} 
\usepackage[protrusion=true,expansion=true]{microtype} 
\usepackage{amsmath,amsfonts,amsthm} 
\usepackage[svgnames]{xcolor} 
\usepackage[format=plain, font=small,labelfont=bf,up,textfont={scriptsize, up}]{caption} 
\usepackage{booktabs} 
\usepackage{fix-cm}	 

\usepackage{sectsty} 

\usepackage{lastpage} 

\DeclareMathSizes{10}{9}{7}{7}
\usepackage{dblfloatfix}    
\usepackage{graphicx}
\usepackage{threeparttable}

\usepackage{lettrine} 
\newcommand{\initial}[1]{ 
\lettrine[lines=3,lhang=0.3,nindent=0em]{
\color{DarkGoldenrod}
{\textsf{#1}}}{}}

\usepackage{titling} 
\usepackage{authblk}

\newcommand{\HorRule}{\color{DarkGoldenrod} \rule{\linewidth}{1pt}} 

\pretitle{\vspace{-30pt} \begin{flushleft} \HorRule \fontsize{22}{22} \usefont{OT1}{phv}{b}{n} \color{DarkRed} \selectfont} 

\title{Protein-protein docking by generalized Fourier transforms on 5D rotational manifolds} 

\posttitle{\par\end{flushleft}\vskip 0.5em} 

\preauthor{\begin{flushleft}\large \lineskip 0.5em } 

\author[1,2]{Dzmitry Padhorny}
\author[2]{Andrey Kazennov}
\author[3]{Brandon S. Zerbe}
\author[3]{Kathryn Porter}
\author[3]{Bing Xia}
\author[3]{Scott E. Mottarella}
\author[4,2,5]{Yaroslav Kholodov}
\author[6]{David W. Ritchie}
\author[3]{Sandor Vajda} 
\author[1,7,8]{Dima Kozakov}

\affil[1]{Department of Applied Mathematics and Statistics, Stony Brook University, Stony Brook, New York 11794, United States}
\affil[2]{Moscow Institute of Physics and Technology, 9 Institutskiy per., Moscow region, Russia, 141700}
\affil[3]{Department of Biomedical Engineering, Boston University, Boston, Massachusetts 02215, United States}
\affil[4]{Innopolis University, 1 Universitetskaya st., Innopolis, Tatarstan, Russia, 420500} 
\affil[5]{Institute of Computer Aided Design of the Russian Academy of Sciences, 19/18, 2-nd Brestskaya st., Moscow, Russia, 123056}
\affil[6]{Inria Nancy, Grand Est, 615 Rue du Jardin Botanique, 54600, Villers-les-Nancy, France}
\affil[7]{Laufer Center for Physical and Quantitative Biology, Stony Brook University, Stony Brook, New York 11794, United States}
\affil[8]{Institute for Advanced Computational Sciences, Stony Brook University, Stony Brook, New York 11794, United States}
\postauthor{\par\end{flushleft}\HorRule} 

\date{} 

\begin{document}

\maketitle 



\initial{E}\textbf{nergy evaluation using fast Fourier transforms enables sampling billions of putative complex structures and hence revolutionized rigid protein-protein docking. However, in current methods efficient acceleration is achieved only in either the translational or the rotational subspace. Developing an efficient and accurate docking method that expands FFT based sampling to 5 rotational coordinates is an extensively studied but still unsolved problem. The algorithm presented here retains the accuracy of earlier methods but yields at least tenfold speedup. The improvement is due to two innovations. First, the search space is treated as the product manifold $\mathbf{SO(3)x(SO(3)\setminus S^1)}$, where $\mathbf{SO(3)}$ is the rotation group representing the space of the rotating ligand, and $\mathbf{(SO(3)\setminus S^1)}$ is the space spanned by the two Euler angles that define the orientation of the vector from the center of the fixed receptor toward the center of the ligand. This representation enables the use of 
efficient FFT methods developed for $\mathbf{SO(3)}$. Second, we select the centers of highly populated clusters of docked structures, rather than the lowest energy conformations, as predictions of the complex, and hence there is no need for very high accuracy in energy evaluation. Therefore it is sufficient to use a limited number of spherical basis functions in the Fourier space, which increases the efficiency of sampling while retaining the accuracy of docking results. A major advantage of the method is that, in contrast to classical approaches, increasing the number of correlation function terms is computationally inexpensive, which enables using complex energy functions for scoring.}

\section{Introduction}
Determining putative protein-protein interactions using genome-wide proteomics studies is a major step toward elucidating the molecular basis of cellular functions. Understanding the atomic details of these interactions, however, requires further biochemical and structural information. While the most complete structural characterization is provided by X-ray crystallography, solving the structures of protein-protein complexes is frequently very difficult. Thus, it is desirable to develop computational docking methods that, starting from the coordinates of two unbound component molecules defined as receptor and ligand, respectively, are capable of providing a model of acceptable accuracy for the bound receptor-ligand complex \cite{Ritchie2008a, Andrusier2008, Vajda2009, Smith2002}. In view of the large number of putative protein-protein interactions, the computational efficiency of docking is also a concern. 

Most global docking methods start with rigid body search that assumes only moderate conformational change upon the association, accounted for by using a smooth scoring function that allows for some level of steric overlaps \cite{Vajda2009}. Rigid docking was revolutionized by the fast Fourier transform (FFT) correlation approach, introduced in 1992 by Katchalski-Katzir and coworkers \cite{Katchalski-Katzir1992}. The major requirement of the method is to express the interaction energy in each receptor-ligand orientation as a sum of $P$ correlation functions, i.e., in the form

\begin{equation}\label{eq:CartMasterIntro}
  \begin{split}
  &E(\alpha,\beta,\gamma, \lambda, \mu, \nu) = \\
  &=\sum_{p=1}^{P} \int \overline{ R_{p}(x, y, z)} \hat{T}(\lambda, \mu, \nu) \hat{D}(\alpha,\beta,\gamma) L_{p}(x, y, z) dV,
\end{split}
\end{equation} 

where $R_p$ and $L_p$ are defined on the receptor and ligand, respectively, $\hat{T}$ and $\hat{D}$ denote translational and rotational operators, and $\alpha,\beta,\gamma$ and $\lambda, \mu, \nu$ are the rotational and translational coordinates. To illustrate how such functions can be used for docking, consider the very simple case with $P=1$, $R_p =-1$ on a surface layer and $R_p =1$ on the core of the receptor, $L_p = 1$ on the entire ligand, and $R_p  =L_p = 0$ everywhere else. It is clear that this scoring function, which is essentially the one used by Katchalski-Katzir and coworkers \cite{Katchalski-Katzir1992}, reaches its minimum on a conformation in which the ligand maximally overlaps with the surface layer of the receptor, thus providing optimal shape complementarity. In later FFT based methods the scoring function has been expanded to include electrostatic and solvation terms \cite{Gabb1997, Chen2003}, and more recently structure-based interaction potentials \cite{Kozakov2006, 
Mintseris2007}, 
substantially improving the accuracy of docked structures. As mentioned, in all scoring functions the shape complementarity term allows for some overlaps, thereby accounting for the differences between bound and unbound (separately crystallized) structures.

Most FFT based methods \cite{Gabb1997, Mandell2001, Chen2003, Kozakov2006, Vakser1996} define $R_p$ and $L_p$ on grids, and use a 3D Cartesian FFT approach to accelerate the sampling of the translational space. The method is based on the idea that the energy function, given by Eq. 1, can be expressed in terms of the Fourier transforms $r_p$ of $R_p$ and $l_p$ of $L_p$. Since the translational operator applied to $l_p$ in the Fourier space is given by
\begin{equation}\label{eq:CartTransIntro}
  \begin{split}
    T(\lambda, \mu, \nu) l_p(n, m, l) =  e^{-2\pi i / N (n\lambda + l \mu + m \nu)} l_{p}(x, y, z),
  \end{split}
\end{equation}
where $i = \sqrt{-1}$, accounting for the orthonormality of Fourier basis functions and interchanging the order of integration and summation yield
\begin{equation}\label{eq:CartFFTIntro}
  \begin{split}
    &E(\alpha,\beta,\gamma, \lambda, \mu, \nu) =\\
    &=\sum_{p=1}^{P} \sum_{nlm} \overline{r_{p}(n, l, m)} l_{p}(\alpha, \beta, \gamma, n, l, m) e^{-\frac{2\pi i} {N} (n\lambda + l \mu + m \nu)},
  \end{split}
\end{equation}
which is the expression for the inverse Fourier transform of the Fourier images $r_p (n, m, l)$ and $l_p (\alpha,\beta,\gamma, n, l, m)$ as stated by the convolution theorem. Thus, for a given rotation $E$ can be calculated over the entire translational space using $P$ forward and one inverse fast Fourier transforms. If $N$ denotes the size of the grid in each direction, then the efficiency of this approach is $O(N^3 log N^3)$ as compared to $O(N^6)$ when energy evaluations are performed directly. Owing to the high numerical efficiency of the FFT based algorithm it became computationally feasible, for the first time, to systematically explore the conformational space of protein-protein complexes evaluating the energies for billions of conformations, and thus to dock proteins without any a priori information on the expected structure of their complex.

In spite of the usefulness of the above algorithm, using FFTs only in translational space has three major limitations. First, FFTs on a new grid must be computed for each rotational increment of the rotating molecule, thus acceleration applies only to half of the degrees of freedom (Fig. 1). Second, each term in the scoring function requires a separate FFT calculation. Thus, accounting for electrostatics, desolvation, and particularly for pairwise interactions substantially increases the required computational efforts. Third, experimental techniques such as NMR Nuclear Overhauser effect (NOE) measurements and chemical crosslinking yield information on approximate distances between interacting residues across the interface, and this information can be used to perform the docking subject to pairwise distance restraints. Unfortunately each pairwise distance restraint requires a new correlation function term. Since the required computational effort is 
proportional to $P$, the number of correlation functions in the energy expression, the increasing complexity reduces the numerical advantage of the FFT approach.

In principle, the above problems can be avoided by applying the transforms first, and then moving the proteins in the Fourier space without the need for re-computing the transforms. However, it is difficult to carry out rotations in the translational Fourier space, and thus to perform rotations efficiently it is natural to use spherical coordinates. Accordingly, a few groups have developed such docking algorithms \cite{Ritchie2000, Ritchie2008}.  
Most notable is the Hex method of Ritchie and Kemp \cite{Ritchie2000}, 
which represents protein shapes using Fourier series expansions of spherical 
harmonic and Gauss-Laguerre polynomials. This representation allows rotational searches to be accelerated by angular FFTs, 
and it enables translations to be calculated analytically in the Fourier basis \cite{Ritchie2008}. A similar approach has been developed by Chacon's group \cite{Kovacs2003,garzon2009frodock} in which translations are calculated numerically. However, both approaches were found to have lower accuracy than traditional Cartesian 
FFT sampling \cite{Ritchie2008}. 
This may attributed to three main factors.  Firstly, the energy functions used were less detailed than in some of the Cartesian approaches. 
In particular, we used only van der Waals and electrostatic terms \cite{Ritchie2008}. 
Secondly, because the computational cost of the polar Fourier translation matrices grows as $O(N^5)$, 
the polar Fourier representation is limited to using relatively low order expansions, 
which limits the achievable accuracy. Finally, rotational space is not euclidean space, but rather a manifold (i.e space which is locally Euclidean, but globally has more complex structure), which precludes straightforward generalization to  effective five-dimensional (5D) rotational  euclidean FFT approach.  This  observation in conjunction  with high-memory intensiveness  of higher order FFTs, resulted   in practice, that 5D rotational  FFTs  were less effective than simple 1D FFTs \cite{Ritchie2010}.
While we showed previously that the polar representation allows an elegant 5D 
factorisation of multi-term potentials \cite{Ritchie2008}, 
previous efforts to exploit this property have until now had limited success.

In this paper we describe a novel fast manifold   Fourier transform (FMFT) algorithm that eliminates the above shortcomings and, on the average, results in a 10-fold decrease in computing time while retaining the accuracy of the traditional Cartesian FFT based docking. Developing the FMFT  method we took advantage of the  generalization of the Cartesian FFT approach to the rotational group manifold $SO(3)$ by Kostelec and Rockmore \cite{Kostelec2008}. The basis for using this algorithm was recognizing that the 5D rotational search space can be regarded as the product manifold $SO(3)\times(SO(3) \setminus S^1)$, where the rotation group $SO(3)$ represents the space of the rotating ligand and $(SO(3) \setminus S^1)$ is the space spanned by the two Euler angles that define the orientation of the vector from the center of the fixed receptor to the center of the ligand (Fig.\ref{fig:FFT_docking}). This is important, because the algorithm by Kostelec and Rockmore \cite{Kostelec2008} can be easily extended to the $SO(3)\times(SO(3)\setminus S^1)$ manifold. However, as mentioned above, a general shortcoming of using Fourier decomposition in spherical spaces is the relatively slow convergence of the series of spherical basis functions. Thus, using a large number of terms reduces computational efficiency, whereas truncating the series limits the accuracy of the energy values calculated by the method. Therefore, a key factor explaining the success of our 
manifold FFT docking method is that we select the centers of highly populated clusters of docked structures rather than low energy conformations as predictions of the complex. This approach becomes feasible because we 
globally and systematically sample the rotational/translational space of the ligand protein on a grid, and hence we can calculate an approximate partition function of the form $Z = \sum_{j} exp(-E_{j} /RT)$, where $E_{j}$ is the energy of the $j^{th}$ pose, and we sum over all poses. For the $k^{th}$ low energy cluster the partition function is given by $Z_{k} = \sum_{j} exp(-E_{j} /RT)$, where the sum is restricted to poses within the cluster. Based on these values, the probability of the $k_{th}$ cluster is given by $P_{k} = Z_{k}/Z$. However, since the low energy structures are selected from a relatively narrow energy range, and the energy values are calculated with considerable error, it is reasonable to assume that these energies do not differ from each other, i.e., $E_{j}=E$ for all $j$ in the low energy clusters. This simplification implies that $P_{k}=exp(-E/RT)\times (N_{k}/Z)$, and thus the probability $P_{k}$ is proportional to $N_{k}$, where $N_{k}$ is the number of structures in the $k^{th}$  cluster. Therefore we select the centers of highly populated clusters of docked structures, rather than low energy conformations, as predictions of the complex. This approach does not require very accurate energy evaluation, and hence it is sufficient to use a limited number of spherical basis functions in the Fourier space, increasing numerical efficiency without noticeable loss in docking accuracy. 

The high efficiency of the FMFT algorithm enables solving very demanding docking problems, way beyond what was considered feasible in the past. After demonstrating that the accuracy of FMFT is comparable to that of the traditional Cartesian FFT based docking, we present here a few applications that require a large number of docking calculations. Such problems include docking ensembles of models obtained by nuclear magnetic resonance (NMR) or homology modeling, and exploring a large number of putative peptide conformations in peptide-protein docking. As will be described, an additional and very favorable property of the FMFT algorithm is that the required computational efforts are almost completely independent of the number P of the correlation function terms in the energy expression given by eq. 1, and hence the method can be efficiently used with scoring functions of arbitrary complexity. In contrast, in the traditional FFT approach the efforts are proportional to P, and hence it is difficult to perform docking subject to pairwise distance restraints, as each restraint gives rise to an additional term in the scoring function. Using FMFT we demonstrate that this problem can be solved effectively without significant increase in running times.


\section{Results and Discussion}
\subsection{FFT based docking on 5D rotational manifolds}
Here we demonstrate that by taking advantage of the special geometry of the space characterizing molecular movement upon protein-protein association it is possible to construct an extremely efficient FFT-based docking algorithm. We present the basic idea of this algorithm as the generalization of the translational FFT method described in the Introduction. Since we plan to work in the rotational space, we change the Cartesian coordinates to polar coordinates $(x, y, z) \rightarrow (r, \theta, \phi)$, and consider the generalization of the Fourier transform on the sphere:
\begin{equation}\label{eq:ManiForwardFFT}
  \begin{split}
    R(r, \theta, \phi) = \sum_{nlm}^{N} r(n, l, m) R_{nl}(r) d_{lm} (cos{\theta}) e^{-i m \phi}
  \end{split}
\end{equation} 
where $R_{nl}(r)$ are radial basis functions, $r(n,l,m)$ are generalized Fourier coefficents, $d_{lm}(cos \theta)$ are Laguerre polynomials \cite{zare2013angular}, and $N$ is the number of the basis functions used. Eq. \ref{eq:ManiForwardFFT} looks like a Fourier transform but $e^{-i m \phi}$ is replaced by $d_{lm} (cos \theta) e^{-i m \phi}$ which shows the non-Cartesian properties of the sphere \cite{Driscoll1994}.

\begin{figure*}[h]
\begin{center}
\centerline{\includegraphics[width=.70\textwidth]{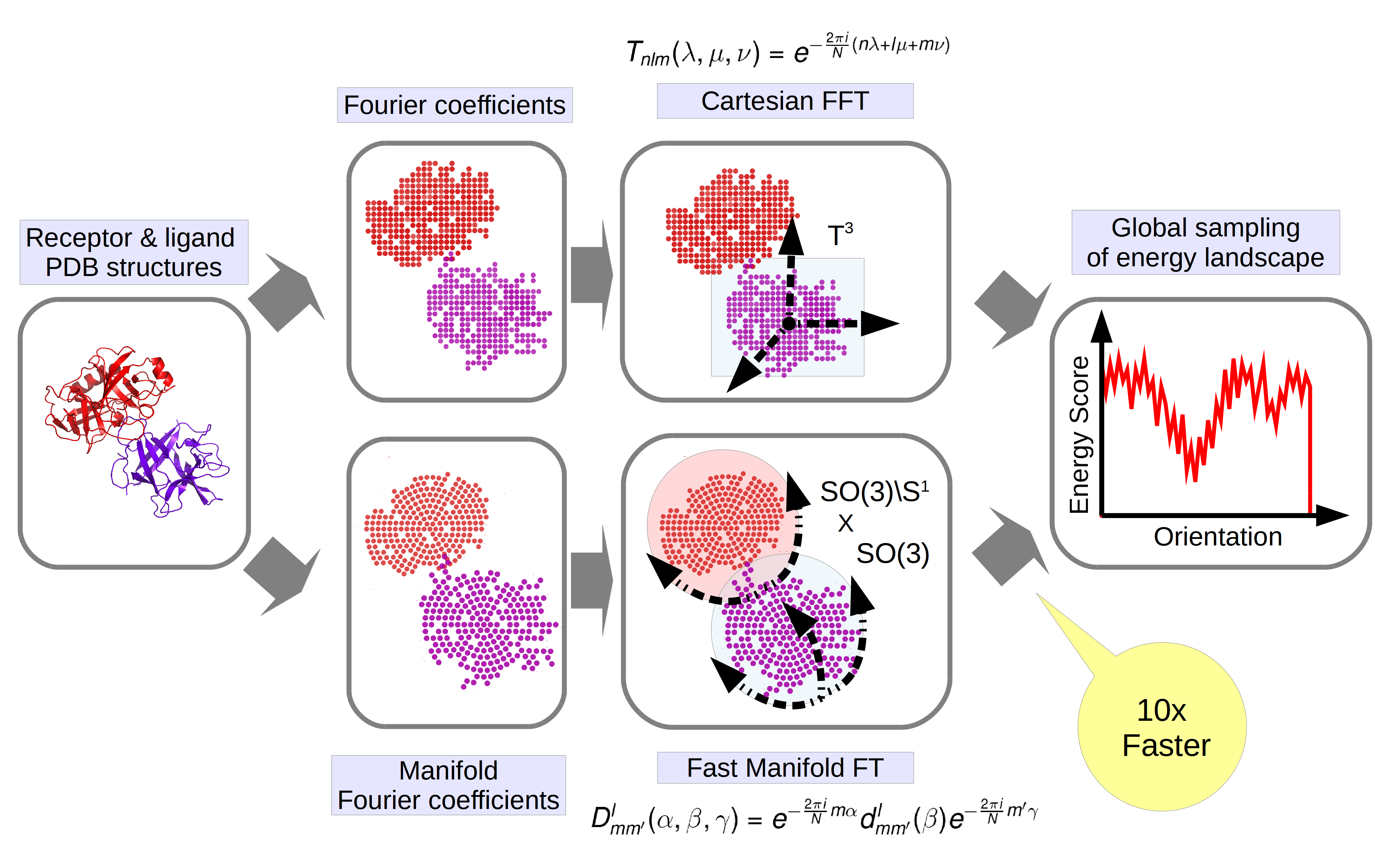}}
\caption{Schematic representation of FFT based docking methods. In Cartesian FFT sampling (upper path) the ligand protein is translated along 3 Cartesian coordinates in Fourier space using the translational operator $T$. The translation must be repeated for each rotation of the ligand. In 5D FMFT docking (lower path) the direction of the vector from the center of the receptor to the center of the ligand is defined by 2 Euler angles, and the ligand is rotated around its center, resulting in the search space $S^2 \times SO(3)$. All rotations are performed in generalized Fourier space, where $D$ denotes the rotational operator. Note that we actually sample $(SO(3)\setminus S) \times SO(3)$ rather than $S^2 \times SO(3)$, which can be done by a straightforward extension of the efficient FFT method developed for the rotational group $SO(3)$. The only traditional search is the one-dimensional translation along the vector between the centers of the two proteins. }\label{fig:FFT_docking}
\end{center}
\end{figure*}

Consider again the derivation of the convolution theorem (Eq. \ref{eq:CartMasterIntro}), but this time on the manifold $(SO(3)\setminus S) \times SO(3)$ shown in the lower path of Fig.\ref{fig:FFT_docking}. The translation of the ligand can be represented as the rotation of the receptor, followed by the translation of the ligand along the z axis,
\begin{equation}\label{eq:ManiMaster}
  \begin{split}
    E(z,\beta,\gamma,\alpha',\beta',\gamma') &=\\
    &=\sum_{p=1}^{P} \int \hat{T}(-z) \hat{D}(0,\beta,\gamma) R_{p}(\rho,\theta,\phi)\\ & \hspace{30pt} \times \hat{D}(\alpha',\beta',\gamma') L_{p}(\rho,\theta,\phi) dV
  \end{split}
\end{equation}
Rotations of the receptor can be expressed as follows:
\begin{equation}\label{eq:ManiRot1}
  \begin{split}
    &D(\alpha,\beta,\gamma) R(r,\theta, \phi) = \\
    &= \sum_{nlm} R_{nl}(r) Y_{lm}(\theta, \phi) \sum_{m_1} D_{mm_1}^{l}(\alpha,\beta,\gamma) r(n,l,m_1),
  \end{split}
\end{equation}
where $Y_{lm}(\theta, \phi)$ denotes spherical harmonics, and 
\begin{equation}\label{eq:ManiRot2}
D_{mm'}^{l}(\alpha, \beta, \gamma) = e^{-im\alpha} d_{mm'}^{l}(\beta) e^{-im'\gamma}
\end{equation}
are Wigner rotation matrices with $d^{l}_{mm1} (\beta)$ denoting Wigner d-functions, related to Jacobi polynomials \cite{Kostelec2008}.
Eqs. \ref{eq:ManiRot1} and \ref{eq:ManiRot2} show that the rotational operator in the rotational group $SO(3)$ acts on generalized Fourier coefficients the same way as the translation operator acts on 
Fourier coefficients in the Cartesian space (see Eq. 2), apart from the asymmetry of the middle angle $\beta$ which requires special treatment. Describing the translation of the ligand along the $z$ axis in the Fourier space is far from simple, and requires updating a set of coefficients. However, it is only one degree of freedom (as opposed to 3 degrees in the Cartesian space), and hence it can be accomplished relatively efficiently \cite{Ritchie2005}. Now we apply the translation operator and the rotation operator (Eq. \ref{eq:ManiRot2}) to the integral in Eq. \ref{eq:ManiMaster}. Based on the orthonormality of the generalized Fourier basis functions, interchanging the order of integration and summation yields
\begin{equation}\label{eq:ManiFFT}
  \begin{split}
  &E(z,\beta,-\gamma,\alpha',\beta',\gamma') =\\ 
  &=\sum_{\stackrel{l l_1}{m m_1 m_2}} \Big( \sum_{n n_1} \sum_p \overline{r_p(n_1, l_1, m_1)} l_p(n, l, m_2) T_{n l n_1 l_1}^{|m|}(z) \Big) \\
  & \hspace{50pt} \times d^{l_1}_{m m_1} (\beta) d^{l}_{m m_2} (\beta') e^{-i(m\alpha' + m_1\gamma + m_2\gamma')}
  \end{split}
\end{equation}
Note that Eq. 8 is similar to Eq. \ref{eq:CartFFTIntro} in Cartesian coordinates, with the difference that instead of a 3D inverse Fourier transform (IFT) we have generalized fast manifold Fourier transform (FMFT), which involves the Wigner d-functions $d^{l}_{mm1} (\beta)$. However, the really important difference is in the order of the transforms and the summation of correlation functions. In Eq. 3, for each rotation of the ligand, we have to calculate the Fourier transforms $l_p (\alpha,\beta,\gamma, n, l, m)$ for each of the $P$ components of the ligand energy function separately, form the product with the transform $r_p (n, m, l)$ of the $p$th component of the receptor energy function, sum all terms, and take the inverse transform. In contrast, according to Eq. 8 we calculate the sum of initial pre-calculated generalized Fourier coefficents in the internal loop only once, and perform all rotations in Fourier space rather than calculating an FFT for each rotation. This allows us to calculate multiple 
energy terms using a single  fast Manifold Fourier transform (FMFT) for each translation. Since inverse manifold Fourier transforms can be efficiently calculated by methods due to \cite{Kostelec2008}, this new approach provides substantial computational advantage, particularly if the number $P$ of the correlation functions in the energy expression is high.

\subsection{Execution times}
Execution times of the FMFT sampling algorithm were measured by docking unbound structures of component proteins in 51 enzyme-inhibitor pairs from the established Protein Docking Benchmark \cite{Hwang2010} (see Table S1). The times were compared to those required for docking the same proteins using PIPER, a protein docking program based on the Cartesian FFT approach \cite{Kozakov2006}. Using the 
FMFT algorithm the average execution time was 15.39 min. In comparison, the average execution time for the same set of proteins using PIPER was 232.15 min, indicating that FMFT speeds up the calculations approximately 15-fold. Using parallel versions of the algorithms on 16 CPU cores, the average execution times measured were 2.67 and 20.19 minutes for FMFT and PIPER, respectively, which shows about 7.5-fold speedup. In all cases Intel Xeon E5-2680 processors were used to run the programs. 

\subsection{Application 1: Constructing enzyme-inhibitor complexes} 
The quality of results was also evaluated by docking the 51 enzyme-inhibitor pairs using both FMFT and PIPER (Table S2). As mentioned, model selection in PIPER involves a clustering procedure in which low energy poses generated by the sampling are clustered and cluster centers are reported as final models with ranks assigned according to populations of these clusters. We used the same approach when using the FMFT algorithm for sampling. In addition, the scoring function was the same one normally used by PIPER for docking enzyme-inhibitor pairs.
\begin{figure*}[b]
\begin{center}
\centerline{\includegraphics[width=.70\textwidth]{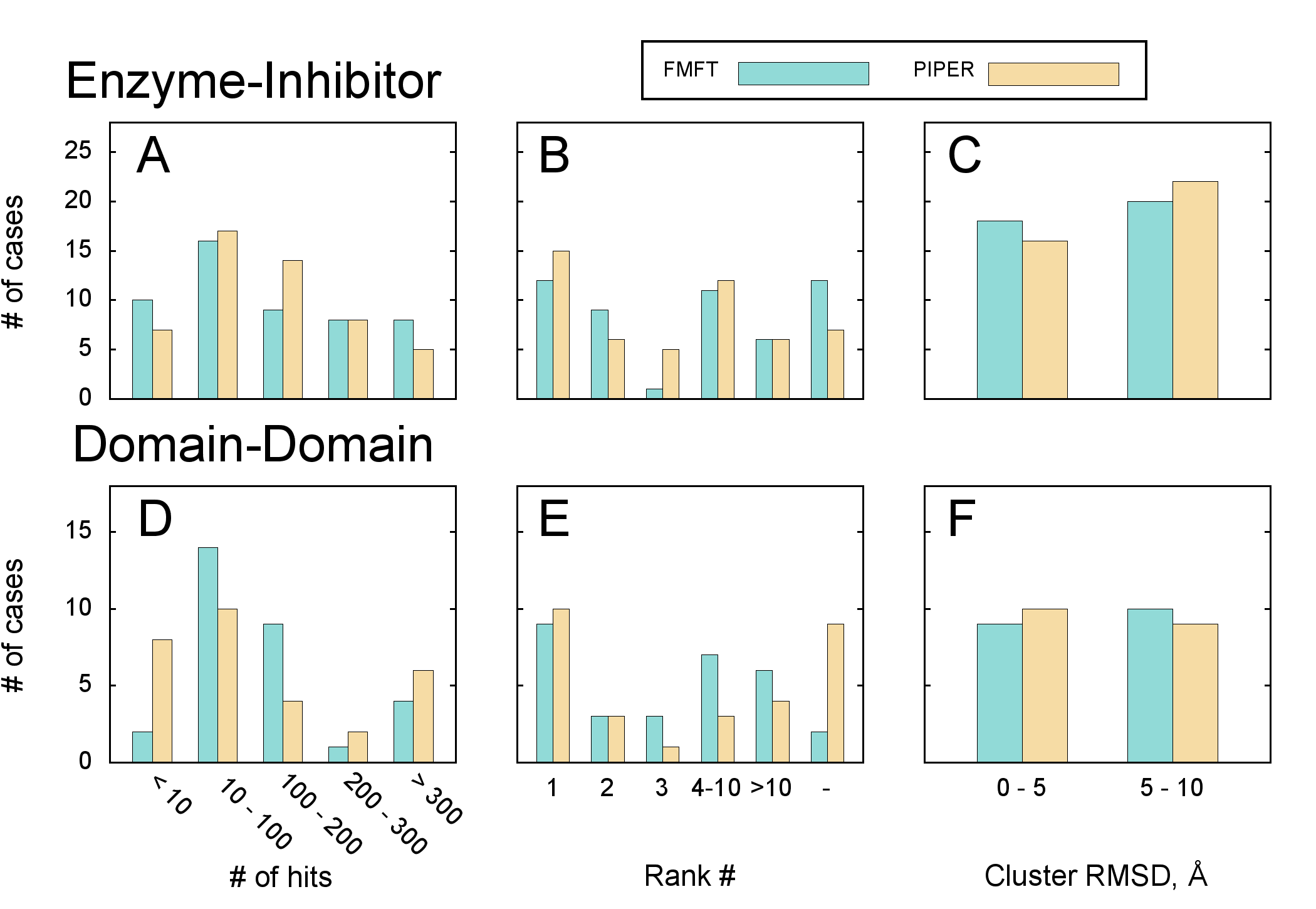}}
\caption{Results of docking enzyme-inhibitor and domain-domain pairs. Bar heights represent the number of docking cases that fall into an appropriate category. (A) The number of hits among the 1000 low energy poses generated for enzyme-inhibitor complexes. (B) Ranking of final near-native models for enzyme-inhibitor complexes. (C) $C_{\alpha}$ IRMSD of the final model for enzyme-inhibitor complexes (here only cases with both FMFT and PIPER producing a near-native model were taken into account).  (D) The number of hits among the 1500 low energy poses generated for domain-domain complexes. (E) Ranking of final near-native models for domain-domain complexes. (F) $C_{\alpha}$ IRMSD of the final model for domain-domain complexes. As in (C), only cases with both FMFT and PIPER producing a near-native model were taken into account).   }\label{fig:dockingResults}
\end{center}
\end{figure*}
Figures 2A,2B, and 2C show the results of docking. The number of hits shown in Fig.\ref{fig:dockingResults} A is the number of near-native poses, defined as having less the 10\AA\:  $C_{\alpha}$ interface RMSD (IRMSD) from the native complex, generated by each of the two algorithms. Note that IRMSD is calculated for the backbone atoms of the ligand that are within 10\AA\: of any receptor atom after superimposing the receptors in the X-ray and docked complex structures. We found that the number of poses with less than 10\AA\: IRMSD is a good measure of the quality of sampling the energy landscape in the vicinity of the native structure. Figures 2B and 2C show the properties of models obtained by clustering low energy poses using pairwise IRMSD as a distance metric. A large number of low energy poses typically yields a well-populated and thus highly ranked near-native cluster, reported as one of the final models. Based on all these results, FMFT and PIPER show comparable docking performance, both in terms of the number of 
near-native structures (Fig.\ref{fig:dockingResults}A), the ranks of the clusters that define the final near-native models (Fig.\ref{fig:dockingResults}B), and the IRMSD (Fig.\ref{fig:dockingResults}C) of these models.

\subsection{Application 2: Docking interacting protein domains} 
We further compared FMFT and PIPER by docking interacting domains extracted from proteins that are defined as "other" type in the Protein Docking Benchmark \cite{Hwang2010} (Tables S3 and S4). This problem is generally more challenging than docking inhibitors to enzymes since the "other" category includes complexes with highly variable properties. Restricting consideration to individual domains eliminates the additional problem that the domains in multidomain proteins may shift relative to each other, affecting the docking results. Similarly to the results obtained for enzyme-inhibitor complexes, FMFT and PIPER show comparable performance (see Figures 2D, 2E, and 2F, and Table S3). Although PIPER generates large numbers ($>200$) of near-native structures for more complexes than FMFT, the number of complexes with very few ($<10$) such near-native structures is substantially less using FMFT than using PIPER. Thus, FMFT shows better performance for the more difficult-to-dock complexes (see Fig. 2D). In addition,
 using PIPER the number of models that are not ranked in the top 10 is much higher than using FMFT (See Fig. 2E). Based on these results, FMFT performs as well as PIPER.

\begin{figure*}[b]
\begin{center}
\centerline{\includegraphics[width=.95\textwidth]{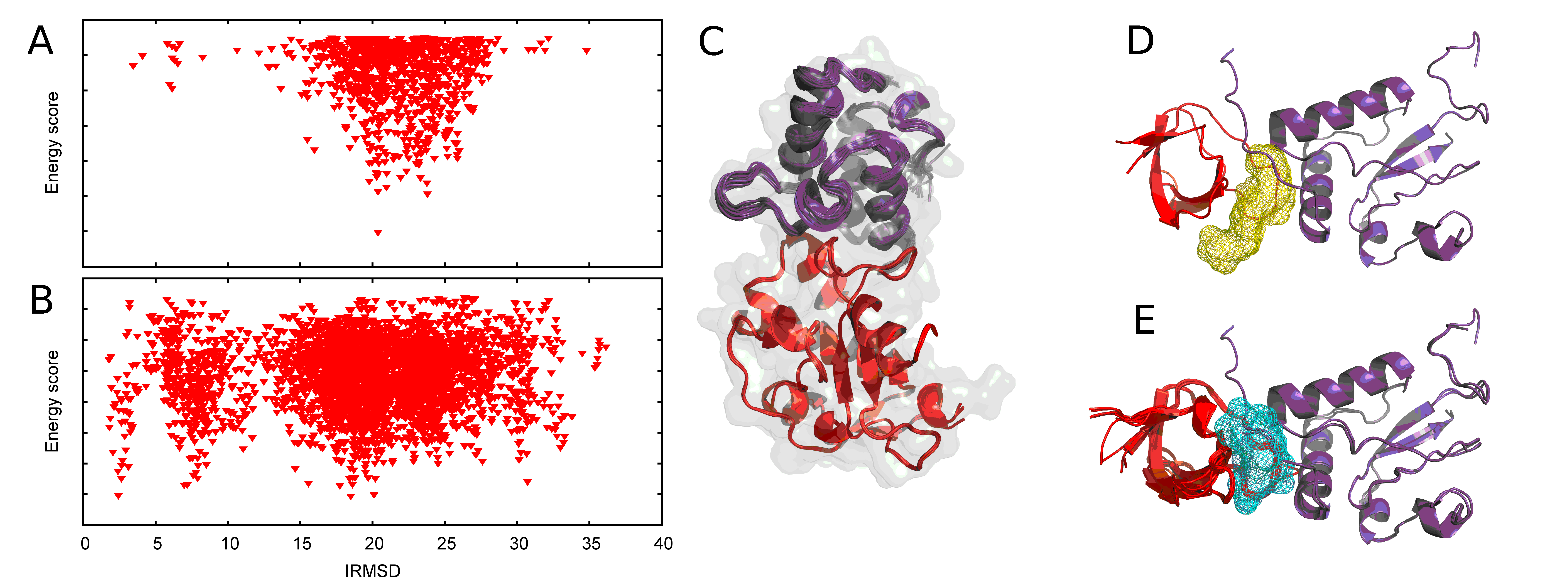}}
 \caption{Docking of structural ensembles. (A) Sampling the interaction energy landscape using a single E9 DNase domain structure and the first NMR model of IM9. The docking does not capture any near-native energy minimum. (B) Consensus energy values from the 80 pairwise dockings of four different X-ray structures of the E9 DNase domain to 20 NMR models of the IM9 protein. (C) Cartoon representation of the four E9 DNase domain and 20 IM9 structures used for docking, superimposed on the structure of the native complex (gray shade). (D) Binding site identification for the Nef-Fyn(R96I)SH3 complex obtained by docking the highest sequence identity models alone. (E) Using multiple homology models of the receptor and the ligand to identify the binding site for the Nef-Fyn(R96I) SH3 complex results into a more specific prediction.}\label{fig:bsites_nmr}. 
\end{center}
\end{figure*}
\subsection{Application 3: Accounting for pairwise distance restraints}
An important consideration for selecting a docking method is the maximum complexity of the scoring function that still allows for solving problems with reasonable execution times. As mentioned, all FFT-based approaches require the use of scoring functions that can be written as sums of correlation functions. This is not a major limitation, since such functions may include many commonly used physics-based energy terms, such as steric repulsion, van der Waals interaction, and Coulombic electrostatics. It has also been shown that some energy terms that are not inherently correlation-based, such as the widely used pairwise interaction potentials, can be efficiently approximated by a sum of several correlation functions \cite{Chuang2008}. Altogether, this makes the number of correlations a crucial parameter, since this number effectively defines the complexity of the scoring function in the particular sampling run.

\begin{figure}[t]
\begin{center}
\centerline{\includegraphics[width=.4\textwidth]{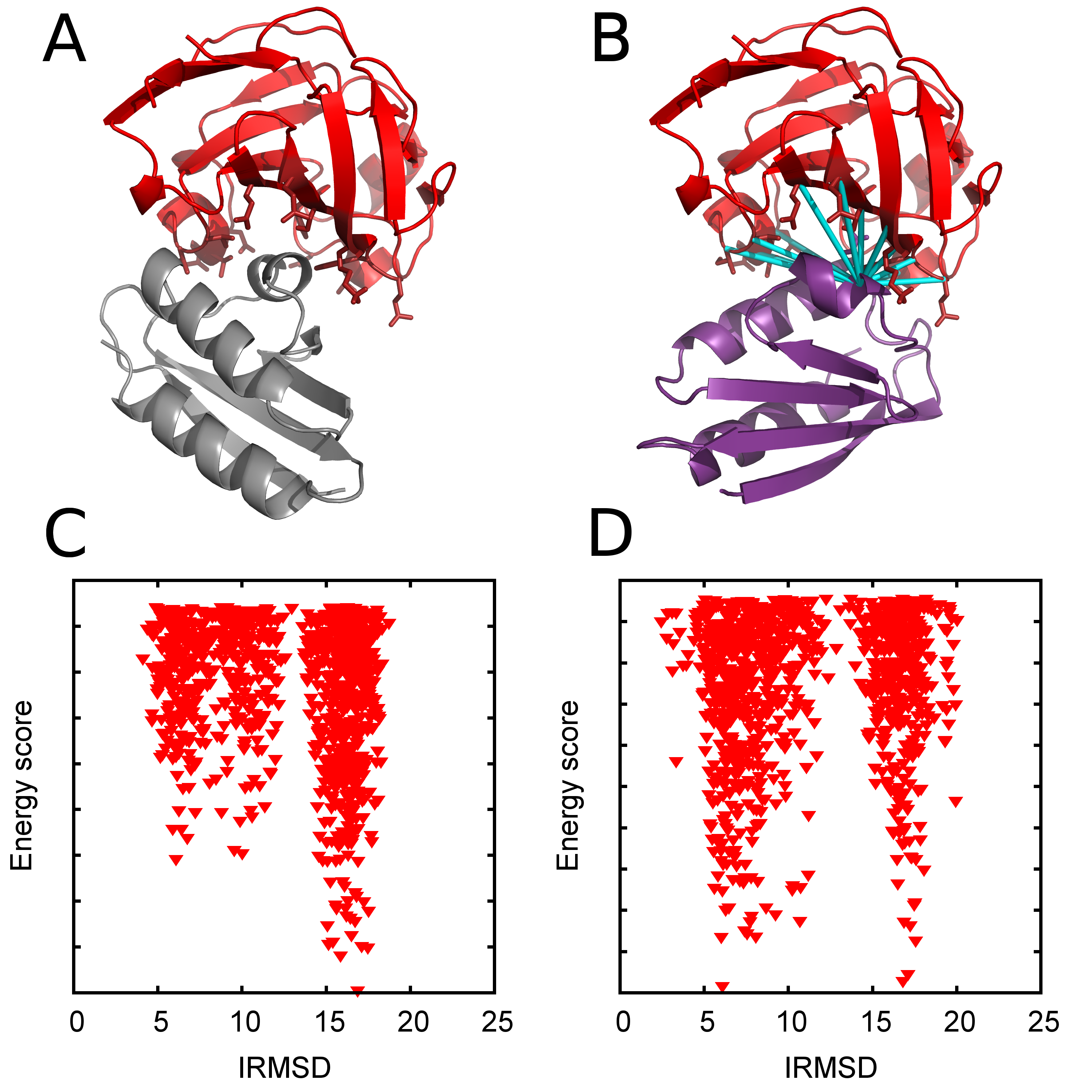}}
\caption{Docking of E2A and HPR proteins.  (A) Model defined by the most populated cluster obtained without restraints. (B) Model defined by the most populated cluster obtained with restraints. A set of cyan cylinders represents one of the 20 restraints. (C) IRMSD versus energy score for docking without restraints. (D) IRMSD versus energy score for docking with restraints. Incorporation of experimental restraints substantially increased the population of the near-native cluster.}\label{fig:constrResults}
\end{center}
\end{figure}

One important task, especially demanding in terms of scoring function complexity, is incorporating pairwise distance restraints, based on known interactions between residue pairs, into the docking procedure. Such restraints can be derived in a variety of experiments, including NMR, cross-linking and mutagenesis assays \cite{Dominguez2003}. The restraints can be implemented as short-distance attractive terms in the scoring function, but each will add a correlation function term. As mentioned, in Cartesian FFT the number of transforms required is proportional to the number $P$ of correlation functions (see Eq. 3), whereas in FMFT the number of transforms is independent of $P$. To demonstrate this difference we considered calculating the structure of the E2A-HPR complex \cite{Garrett1997} using a set of ambiguous interaction restraints (AIRs), based on NMR titration data \cite{Dominguez2003}. Each restraint is specified as a residue in one of the proteins, and a set of residues on the partner protein that are 
in contact with the first residue, where "contact" means $\leq$ 3 \AA \ distance between any two atoms of the residue pair. Docking was performed using both FMFT and PIPER. Incorporation of restraints increased the population of the near-native cluster from 201 to 410, which became the most populated cluster and thus provided the putative model of the complex (see Fig. 3 and Table S5) without any significant change in the IRMSD of the cluster center (5.25 \AA\: for the unrestrained case versus 5.15 \AA\: for the restrained). Adding the restraints increased the number of correlations function terms in the scoring function from 8 to 28. For PIPER this resulted in a proportional increase in execution time (from 96.15 to 373.80 minutes). In contrast, running FMFT the execution time barely changed, from 12.32 minutes to 15.30 minutes. This result demonstrates that FMFT can be used with very complex scoring functions (see Fig. S2).

\subsection{Application 4: Docking ensembles of NMR models}
Multiple docking runs may be required when one or both component proteins are given as ensembles of structures, obtained by nuclear magnetic resonance (NMR) experiments or by extracting snapshots from molecular dynamics simulations. Accounting for multiple structures may substantially improve docking results. As an example, we considered calculating the complex formed by the E. Coli Colicin E9 DNase domain and its cognate immunity protein IM9. Four different X-ray structures of the unbound E9 DNase domain were docked in a pairwise manner to 20 NMR models of the IM9 protein, thus performing 80 docking calculations. Figures 4A, 4B, and 4C show the docking results. In short, merging the 50 lowest energy poses from each docking run, followed by clustering, provided a 2.94 \AA\: IRMSD model of the complex ranked 5th. In contrast, docking a single receptor structure with the first NMR model of the ligand in the ensemble using a standard docking protocol the best near-native model was ranked 13, and had the IRMSD 
value of 3.45 \AA.

\subsection{Application 5: Identification of binding sites by docking homology models}
It has been shown that protein-protein interaction sites can be found by determining the highly populated interfaces in the ensemble of structures generated by global docking \cite{Hwang2014, Fernandez-Recio2004}. We implemented this approach by clustering the ``interfacial'' atoms in the low energy docked poses. While this method usually requires structures of the component proteins, we extended the approach to proteins with yet undetermined structures by docking multiple homology models. The extended method was applied to determining the interface in the Nef - Fyn(R96I)SH3 complex (PDB entry 1EFN). A total of 10 homology models of the SH3 domain and 2 models of the Fyn(R96I) protein were constructed and docked in pairwise manner (see Table S6 for details), thus requiring 20 docking runs. As shown in Figures 4D and 4E, docking of multiple homology models of the component proteins increased the accuracy of binding site prediction, compared to the result of using the maximum sequence identity models alone. 

\subsection{Application 6: Docking flexible peptides} The difficulty in docking short linear peptides is that their structure in solution is generally unknown and may be ill defined. One possible solution is to dock a variety of peptide conformations, thus requiring multiple docking runs. We have recently developed an 
algorithm based on the use of structural templates extracted from the Protein Data Bank (PDB) with sequences that matched the known sequence motif in the peptide. 
\begin{figure}[h]
\begin{center}
\centerline{\includegraphics[width=.4\textwidth]{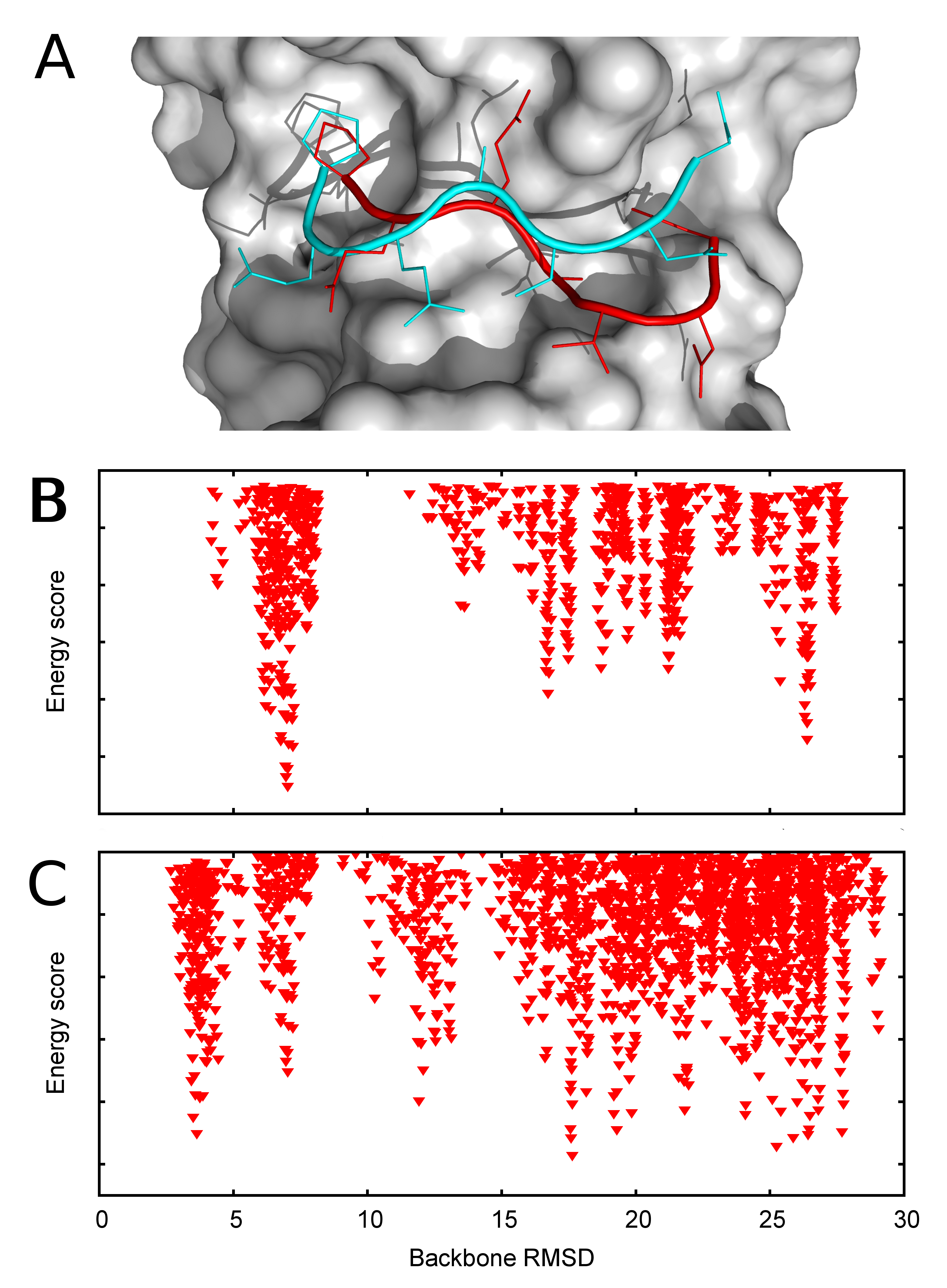}}
\caption{Docking of the ace-PQQATDD peptide to TRAF2. (A) Bound structure of the peptide (red) and the 3.3 \AA\: model, ranked 4th (cyan). (B) Peptide backbone RMSD versus scoring function when docking of the most common structural template alone. (C) Peptide backbone RMSD versus scoring function when using all 25 templates. Docking the ensemble substantially improves the results, and yields  samples with less then 4.0 \AA\: backbone RMSD.}\label{fig:peptides}
\end{center}
\end{figure}
These templates were docked individually using the FMFT algorithm. From each run 250 low energy poses were retained, the pooled peptide structures were clustered, and the highly populated cluster centers were reported as final models as in all applications of our docking algorithm. 

Here we demonstrate this algorithm by docking the ace-PQQATDD peptide to the tumor necrosis factor receptor associated factor 2 (TRAF2). A set of 25 structural templates was used. All templates were docked to the unbound structure of the receptor. A near-native model of the protein-peptide complex was ranked 4th and had the backbone RMSD of 3.3 \AA\: from the conformation in the X-ray structure (see Fig. \ref{fig:peptides} A). Note that docking only the most frequently occurring structural template provides less accurate models as demonstrated in Figures 5B and 5C.  

\section{Conclusions}
Extending the classical 3D Cartesian implementation of the FFT correlation approach to perform rotations in Fourier space without the need for recalculating the transforms has been a long outstanding and extensively studied problem. The main difficulty in developing such methods is that in order to achieve numerical efficiency one can use only a moderate number of spherical basis functions to span the search space, and this may reduce the accuracy of energy evaluation. However, since we base model selection on the population of low energy clusters rather than on energy values, minor deviations in energy generally do not affect the accuracy of final models. Here we present an elegant manifold FFT implementation of 5D search that is more than tenfold faster than the traditional 3D approach. A major advantage of the new method is that the  adding correlation function terms in the scoring function is computationally inexpensive, and hence the method works efficiently with very 
complex energy evaluation models, possibly including pairwise distance restraints that are difficult to deal with in traditional FFT based docking. The improved efficiency implies that we can solve new classes of docking problems, including the docking of large ensembles of proteins rather than just a single protein pair, docking homology models and flexible peptides that may have a large number of potential conformations. We note that the beta version of a code implementing the FMFT algorithm can be downloaded from  \verb|https://bitbucket.org/abcgroup_midas/fmft_dock/|, thus providing opportunity for testing and using the method. 

\section{Methods}
\subsection{FFT sampling on manifolds}
{\footnotesize This section summarizes the implementation of the FMFT approach. For the mathematical details of the algorithm see Supporting Methods. 

The procedure starts with receptor and ligand-associated components of each correlation term of the energy function being represented as sets of coefficients  $\scriptstyle r(n, l, m)$, $\scriptstyle l(n, l, m)$ that appear in the expansion shown as Eq. \ref{eq:ManiForwardFFT}. Here $\scriptstyle 1 \leq n \leq N$, $\scriptstyle 0 \leq l \leq n-1$; $\scriptstyle -l \leq m \leq +m$, where $\scriptstyle N$ governs the order at which the series is truncated. These coefficients, together with the translation range to be sampled (i.e. minimal and maximal distances between protein centers, calculated from the geometrical properties of the proteins), are submitted as input parameters to the program performing the FMFT-based sampling. To improve efficiency, two stages of FMFT sampling are being executed: the first one, performed with a maximal coefficient order $\scriptstyle N = 20$ on a small FFT grid is computationally inexpensive and provides a crude approximation of the energy landscape, which is then used to 
focus the search to the 
translation range potentially containing the energy minima, while the second one is executed with $\scriptstyle N = 30$ on a full-sized FFT grid but performs the sampling only in the refined translation range, thus saving computational resources. 

The actual sampling stage can be described as follows: after loading the input parameters, the program starts to iterate the allowed translation range in steps of 1 \AA. For each translation step, the $\scriptstyle \sum_{n n_1} \sum_p  r_p(n_1, l_1, m_1) l_p(n, l, m_2) T_{n l n_1 l_1}^{|m|}(z)$ product of coefficients and translation matrix elements is calculated, followed by a manifold FFT, which provides the values of energy score for all receptor-ligand orientations corresponding to a fixed distance between the centers of the two proteins. The resulting samples are located on the $\scriptstyle (\beta, \gamma, \alpha', \beta', \gamma')$ Euler angle grid with dimensions of  30 x 59 x 59 x 30 x 59  (or 16 x 30 x 30 x 16 x 30  for the low-order scan). $\scriptstyle K$ (in the order of 1000 for a typical sampling run) lowest energy samples are retained for each translation step. After the entire translation range is processed, the low-energy samples from individual translation steps are merged and re-sorted by 
energy value to select the $\scriptstyle K$ lowest energy samples that are presented as the final results.

It is important to note here that the sampling of the $\scriptstyle S^2 \times SO(3)$ manifold (in practice probed as $\scriptstyle (SO(3) \setminus S^1 )\times SO(3)$ ), provided by the equispaced sampling of Euler angles, is inherently non-uniform. This becomes a significant problem if one seeks to obtain statistical information about the energy landscape of protein interaction, for example, to construct the partition function of the system. To battle this non-uniformity, a special procedure is employed for the selection of low-energy scores. Specifically, once the 5D array of energy scores for a single translation step is acquired, the program starts selecting lowest-scoring conformations and excluding the samples corresponding to the surrounding region from further consideration. Here the "surrounding region" is defined as the subset of elements \{\(\scriptstyle (x, y) \vert x(\beta, \gamma) \subset S^2, y(\beta, \gamma, \alpha', \beta', \gamma') \subset SO(3)\)\} of the $\scriptstyle S^2 \times SO(3)$ 
manifold, for which $\scriptstyle (dist_{S^2}(x, x_{min}) < \varDelta) \land (dist_{SO(3)}(y, y_{min}) < \varDelta)$, where $\scriptstyle \varDelta$ is a cutoff parameter chosen to be 6.0 degrees, which is
slightly less then the grid step of $\scriptstyle 360^{\circ} / 59 = 6.1^{\circ}$. This procedure ensures that the sampling explores substantial fraction of the conformational space rather than producing structures very close to each other. 
}

\section{Acknowledgments}
This work was supported by  NSF AF 1527292  and  NSF DBI 1458509  from the National Science Foundation, grants  NIH R01GM061867 and NIH R01 GM093147  from the National Institute of General Medical Sciences, Russian Scientific Foundation Grant No 14-11-00877 and  US Israel BSF 2009418

\end{document}